\documentclass[prb,superscriptaddress,showpacs,eqsecnum,amsmath,twocolumn]{revtex4}
\usepackage{bm}
\usepackage{graphicx}
\usepackage{math}

\begin{document}

\preprint{WU--HEP--02--07}
\title{Modified Fowler-Nordheim Field-Emission Formulae from a Nonplanar-Emitter Model}
\affiliation{Department of Physics, Waseda University, Tokyo 169--8555, Japan}
\affiliation{Department of Applied Physics, Waseda University, Tokyo 169--8555, Japan}
\affiliation{Kagami Memorial Laboratory for Materials Science and Technology, Waseda University, Tokyo 169--0051, Japan}
\affiliation{Advanced Research Institute for Science and Engineering, Waseda University, Tokyo 169--8555, Japan}
\author{Kazuya Yuasa}
\email{yuasa@hep.phys.waseda.ac.jp}
\author{Ayumi Shimoi}
\affiliation{Department of Physics, Waseda University, Tokyo 169--8555, Japan}
\author{Ichiro Ohba}
\email{ohba@waseda.jp}
\affiliation{Department of Physics, Waseda University, Tokyo 169--8555, Japan}
\affiliation{Kagami Memorial Laboratory for Materials Science and Technology, Waseda University, Tokyo 169--0051, Japan}
\affiliation{Advanced Research Institute for Science and Engineering, Waseda University, Tokyo 169--8555, Japan}
\author{Chuhei Oshima}
\email{coshima@waseda.jp}
\affiliation{Department of Applied Physics, Waseda University, Tokyo 169--8555, Japan}
\affiliation{Kagami Memorial Laboratory for Materials Science and Technology, Waseda University, Tokyo 169--0051, Japan}
\date[]{November 25, 2003}
\begin{abstract}
Field emission formulae, current--voltage characteristics and energy distribution of emitted electrons, are derived analytically for a \textit{nonplanar} (hyperboloidal) metallic emitter model.
The traditional Fowler--Nordheim formulae, which are derived from a \textit{planar} emitter model, are modified, and the assumption of the planar emitter in the F--N model is reconsidered.
Our analytical calculation also reveals the backgrounds of the previous numerical discussion by He \textit{et~al.}~on the effect of the geometry of emitter on field emission.
The new formulae contain a parameter which characterizes the sharpness of the hyperboloidal emitter, and experimental data of field emissions from clean tungsten emitters and nanotip emitters are analyzed by making use of this feature.
\end{abstract}
\maketitle

\section{Introduction}
The Fowler--Nordheim (F--N) theory~\cite{ref:FowlerNordheim,ref:Nordheim,ref:GoodMueller,ref:Young} is one of the most important theories of electron field emission.
It describes experiments, i.e., current--voltage characteristics of field emission current~\cite{ref:FowlerNordheim,ref:Nordheim,ref:GoodMueller} and energy distribution of field emitted electrons,\cite{ref:Young} quite well.
It should be noted, however, that the surface of emitter is assumed to be \textit{planar} in their model~\cite{ref:FowlerNordheim,ref:Nordheim,ref:GoodMueller,ref:Young} although actual emitters are \textit{not planar} literally, and the famous F--N formula $I/V^2\propto\exp(-A/V)$, for example, is derived under this assumption.

This assumption may be considered to be justified since the emission area on the surface of emitter may be limited to a so small region at the apex of the emitter that the area can be regarded as planar.
There is however no rigorous verification of this expectation.
Furthermore, emitters are becoming sharper and sharper nowadays, e.g., nanotip,\cite{ref:NanotipBinhPRL69,
ref:NanotipOshimaASS182} nanotube,\cite{ref:NanotubeRinzlerSci269
} etc., for which the planar emitter model is inappropriate.
It is hence worthwhile to see how the field emission formulae are modified when the (three dimensional) geometry of emitter is taken into account.

Attempts to incorporate geometrical effects into the F--N theory have already been made by several groups.\cite{ref:CutlerImage,ref:Cutler,ref:Lang,ref:NanotipGohdaPRL87}
He \textit{et~al.}~derived bias fields and image charge potentials exactly for the tips shaped like cone, hyperboloid, paraboloid, and sphere on cone,\cite{ref:CutlerImage} and discussed the effect of the geometry of the emitters on field emission:\cite{ref:Cutler}
They numerically obtained (i)~field emission currents from the apex of those emitters and (ii)~(normal) energy distributions of emitted electrons, and they concluded that (i)~the current--voltage characteristics for the nonplanar (and rather sharp) emitters can be fitted with the relationship $I/V^2\propto\exp(-A/V-B/V^2)$, which is different from the one by the F--N model, $I/V^2\propto\exp(-A/V)$, and that (ii)~the energy distributions for the nonplanar models are much wider than that for the F--N planar model.

In this article, we successfully derive \textit{analytical formulae} for current--voltage characteristics and energy distribution for a \textit{hyperboloidal} emitter model (although the image charge effect is neglected).
Even though our formulae are valid only for conventional emitters whose tip radii of curvature are of the order of $100\,\nm$, they still reflect the geometry of the emitter and the traditional F--N formulae are modified.
Furthermore, our analytical calculation enables us to discuss the assumption of the planar emitter in the F--N model, and also helps us understand the backgrounds of the above-mentioned numerical discussion by He \textit{et~al.}
And at the final part of this article, experimental data of field emissions from nanotip emitters~\cite{ref:NanotipOshimaASS182} are analyzed by making use of a parameter, which characterizes the sharpness of the hyperboloidal emitter and never comes in the ordinary F--N theory based on the planar emitter model.

\section{Hyperboloidal Emitter Model}
\begin{figure}[b]
\includegraphics[width=0.45\textwidth]{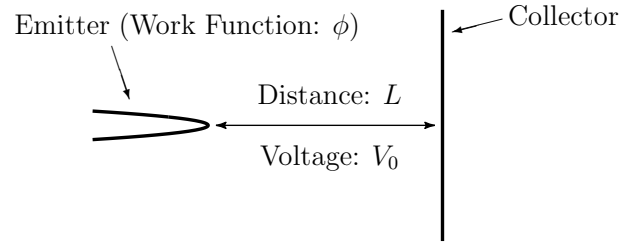}
\caption{A model composed of a hyperboloidal metallic emitter and a planar collector. A typical set of parameters in an actual experiment~\cite{ref:NanotipOshimaASS182} is $L\sim5\,\cm$, $V_0\sim3\,\kV$, and $\phi\sim4.4\,\eV$ (tungsten). The work function of the emitter, $\phi$, is assumed to be uniform.}
\label{fig:Model}
\end{figure}
Our model is illustrated in Fig.~\ref{fig:Model}\@.
It is composed of a hyperboloidal metallic emitter and a planar collector.
The work function of the emitter, $\phi$, is assumed to be uniform on the surface of the emitter.

It is implicitly assumed in the ordinary F--N theory~\cite{ref:FowlerNordheim,ref:Nordheim,ref:GoodMueller,ref:Young} that the emission area on the surface of the emitter, from which electrons are emitted, is limited to a very small region at the apex of the emitter and can be regarded as planar.
The surface of the emitter is thus modelled by a plane, and tunnelling of electrons through a one-dimensional potential barrier at the surface is discussed.
In this article, on the contrary, we duly take the shape of the emitter into account and proceed without such an assumption.
The size of the emission area is also what is to be clarified as a result of the following calculation.

\begin{figure}[b]
\begin{tabular}{c@{\quad}c}
\includegraphics[height=0.195\textwidth]{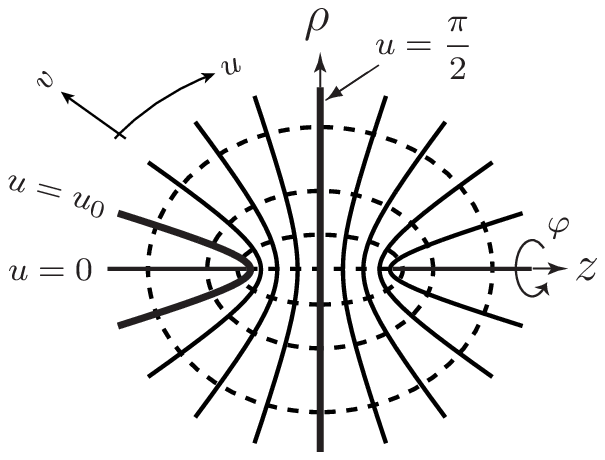}&
\includegraphics[height=0.195\textwidth]{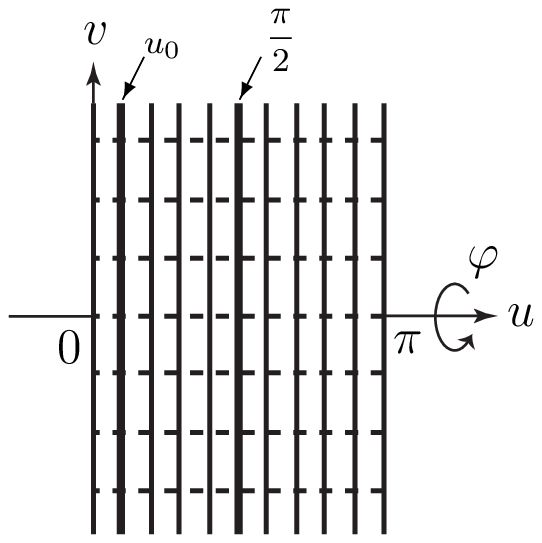}\\
(a)&(b)
\end{tabular}
\caption{An ellipsoidal coordinate system $(u,v,\varphi)$. This is a 3-dim.~orthogonal curvilinear coordinate system. The whole plane in~(a) is mapped onto the region $0\le u\le\pi$ in~(b) through Eq.~(\ref{eqn:EllipsoidalCordinate}) for each $\varphi$.}
\label{fig:Coordinate}
\end{figure}

We work in an orthogonal curvilinear coordinate system, i.e., an ellipsoidal coordinate system $(u,v,\varphi)$ [Fig.\ \ref{fig:Coordinate}(a)].
It is related to the cylindrical coordinate system $(\rho,\varphi,z)$ by
\begin{equation}
\left\{
\begin{array}{l}
\medskip
\displaystyle   z=-a\cos u\cosh v,\\
\displaystyle\rho= a\sin u\sinh v,
\end{array}
\right.
\label{eqn:EllipsoidalCordinate}
\end{equation}
or
\begin{equation}
z+i\rho=-a\cos(u+iv),
\label{eqn:ConformalMapping}
\end{equation}
where the ranges of the coordinates are $0\le u\le\pi$, $0\le v<\infty$, and $0\le\varphi<2\pi$ [Fig.~\ref{fig:Coordinate}(b)].
A surface $v=\text{const.}$~is an ellipsoid, and a surface $u=\text{const.}$~is a hyperboloid.
The expression~(\ref{eqn:ConformalMapping}) with an analytic function shows that this mapping is a conformal one and the ellipsoidal coordinate system in Fig.~\ref{fig:Coordinate}(a) is an orthogonal one.

We attach the ellipsoidal coordinate system in Fig.~\ref{fig:Coordinate}(a) to the model in Fig.~\ref{fig:Model} such that the plane $u=\pi/2$ is on the collector and a hyperboloid $u=u_0$ is on the surface of the emitter.
The parameters $a$ and $u_0$ are related to the distance from the apex of the emitter to the collector, $L$, by
\begin{equation}
L=a\cos u_0
\end{equation}
and to the radius of curvature of the tip, $R$, by
\begin{equation}
R=a\frac{\sin^2\!u_0}{\cos u_0}=L\tan^2\!u_0.
\end{equation}
$u_0$ is the parameter which characterizes the sharpness of the emitter.
Typical values of it are shown in Table~\ref{tab:Curvatures}\@.
\begin{table}
\caption{Radius of curvature $R$ vs $u_0$ for $L=5\,\cm$.}
\begin{ruledtabular}
\begin{tabular}{cccc}
$u_0$      &0.0010&0.0015&0.0020\\\hline
$R$ ($\nm$)&    50&   110&   200
\end{tabular}
\end{ruledtabular}
\label{tab:Curvatures}
\end{table}
It should be noted here that this parameter is not contained in the ordinary F--N theory since the F--N theory is based on the planar emitter model.

In this coordinate system, the Laplacian $\nabla^2$ is given by
\begin{align}
\nabla^2={}&\frac{1}{\sqrt{g}}
\frac{\partial}{\partial u}
a\sin u\sinh v
\frac{\partial}{\partial u}
+\frac{1}{\sqrt{g}}
\frac{\partial}{\partial v}
a\sin u\sinh v
\frac{\partial}{\partial v}\nonumber\\
&{}+\frac{1}{a^2\sin^2\!u\sinh^2\!v}
\frac{\partial^2}{\partial\varphi^2}
\label{eqn:Laplacian}
\end{align}
with $\sqrt{g}=a^3(\sin^2\!u\cosh^2\!v+\cos^2\!u\sinh^2\!v)\sin u\sinh v$, and the Laplace equation $\nabla^2 V=0$ with the boundary conditions $V=0$ on the surface of the emitter and $V=V_0$ on the collector plane, where $V_0$ ($>0$) is the applied voltage, is reduced to
\begin{equation}
\frac{d}{du}\sin u\frac{d}{du}V(u)=0,\quad
\left\{
\begin{array}{l}
\medskip
V(u_0)=0,\\
V(\pi/2)=V_0.
\end{array}
\right.
\end{equation}
It is then easy to obtain the electric potential $V(u)$ between the emitter and the collector:
It reads
\begin{equation}
V(u)=V_0\left(1-\frac{\ln\cot(u/2)}{\ln\cot(u_0/2)}\right).
\label{eqn:ElectricPotential}
\end{equation}
An equipotential surface is a hyperboloid $u=\text{const.}$, and an electric line of force is a curve along an ellipse $v$, $\varphi=\text{const.}$
\begin{figure}[b]
\begin{tabular}{c@{\quad}c}
\includegraphics[height=0.155\textwidth]{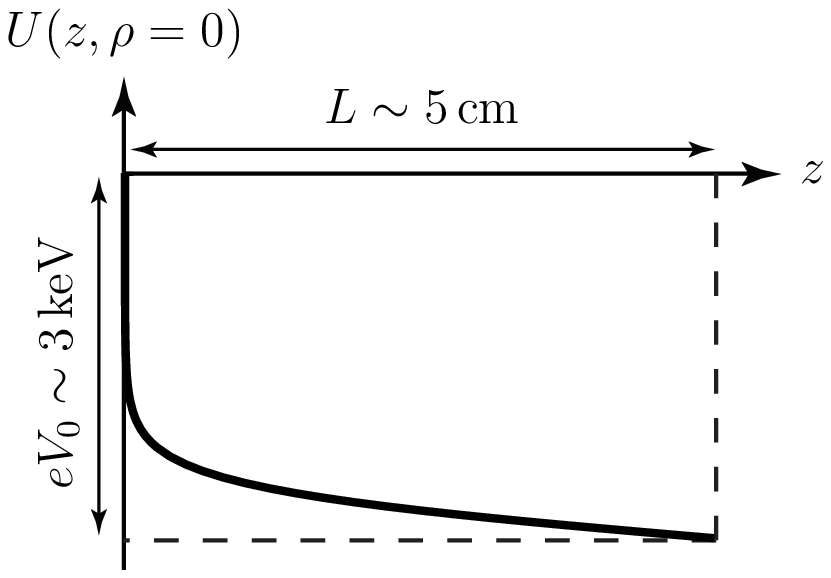}&
\includegraphics[height=0.155\textwidth]{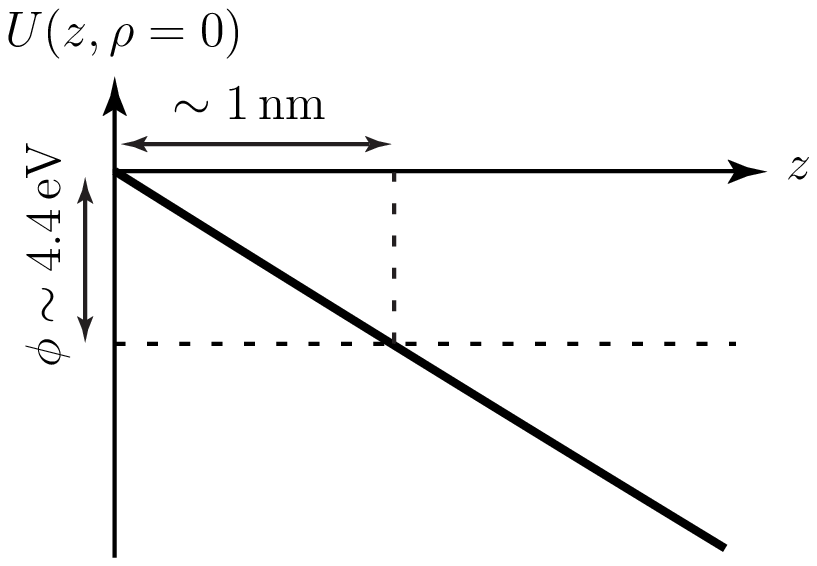}\\
(a)&(b)
\end{tabular}
\caption{Potential energy of an electron, $U$ in Eq.~(\ref{eqn:PotentialEnergy}) with~(\ref{eqn:ElectricPotential}), on the $z$-axis for $L\sim5\,\cm$, $R\sim100\,\nm$, and $V_0\sim3\,\kV$. The $U$-axis is drawn at the apex of the emitter. The potential near the apex in~(a) is enlarged in~(b). (a)~Due to the sharpness of the emitter, a strong field is realized near the apex of the emitter, and (b)~the thickness of the tunnelling barrier is of the order of $1\,\nm$.}
\label{fig:Potential}
\end{figure}
The potential energy of an electron in the electric field,
\begin{equation}
U(u)=-eV(u),
\label{eqn:PotentialEnergy}
\end{equation}
where $-e$ ($<0$) is the charge of the electron, is drawn in Fig.~\ref{fig:Potential} for $\rho=0$.
In this article, we neglect the image charge potential.

\section{Generalization of the F--N Theory for the Hyperboloidal Emitter Model}
Generalizing the F--N theory,\cite{ref:FowlerNordheim,ref:Nordheim,ref:GoodMueller,ref:Young} we calculate the field emission current $I$ for the hyperboloidal emitter model by the equation
\begin{equation}
I=e\rint_0^{2\pi}d\varphi\rint_0^\infty dv\rint_{-\infty}^\infty dE\rint_{-\infty}^EdWN(W,E,v)D(W,v).
\label{eqn:FNEq}
\end{equation}
$E$ and $W$ are the ``total'' and ``normal'' energies of an electron, respectively.\cite{ref:Young}
$N(W,E,v)$ is called supply function,\cite{ref:Young} and $N(W,E,v)\,dWdE\,dv\,d\varphi$ gives the number of electrons incident on the area $dv\,d\varphi$ around the position $(v,\varphi)$ on the face of the potential barrier $u=u_0$ per unit time, with the total energy within the range $E$ to $E+dE$ and the normal energy within the range $W$ to $W+dW$.
$D(W,v)$ is the probability of an electron penetrating the potential barrier at $(v,\varphi)$ with the normal energy $W$.
We obtain the emission current $I$ by summing up the number of \textit{emitted} electrons per unit time, $N(W,E,v)D(W,v)\,dWdE\,dv\,d\varphi$, for all possible energies $(W,E)$ and for all positions $(v,\varphi)$ on the surface of the emitter, i.e., by Eq.~(\ref{eqn:FNEq}).

It should be noted in this formulation that the radius of curvature of the emitter is assumed to be much larger than the de Broglie wave length of an electron ($\sim0.1\,\nm$) since electrons are treated as localized objects. (For conventional emitters, the radii of curvature are of the order of $100\,\nm$.)
Under this assumption, the surface of the potential barrier $u=u_0$ seems planar to an electron, and the other surface of the barrier (the end of the tunnelling region) $u=u_T$, which is separated from the surface $u=u_0$ only by the distance of the order of $1\,\nm$, also seems planar and parallel to the surface $u=u_0$.
This observation leads us to the following explicit formulae for $D(W,v)$ and $N(W,E,v)$.

For the barrier penetration probability $D(W,v)$, let us consider the Schr\"o\-dinger equation
\begin{equation}
\left(-\frac{\hbar^2}{2m}\nabla^2+U(u)\right)\psi(u,v,\varphi)
=E\psi(u,v,\varphi)
\label{eqn:SchroedingerEq}
\end{equation}
with the Laplacian $\nabla^2$ given in Eq.~(\ref{eqn:Laplacian}) and the potential barrier $U(u)$ in Eq.~(\ref{eqn:PotentialEnergy}) with~(\ref{eqn:ElectricPotential}).
($m$ is the mass of an electron.)
Decomposing the wave function as $\psi(u,v,\varphi)=e^{i\varTheta(u,v)/\hbar}e^{ip_\varphi\varphi/\hbar}$, where $\varTheta(u,v)=\varTheta_R(u,v)+i\varTheta_I(u,v)$ is a complex-valued function and $p_\varphi$ a real constant (angular momentum around the $z$-axis), the Schr\"odinger equation~(\ref{eqn:SchroedingerEq}) is reduced in the WKB approximation~\cite{ref:Schiff} to
\begin{align}
&\frac{a\sin u\sinh v}{\sqrt{g}}\,\biggl[
\left(\frac{\partial\varTheta}{\partial u}\right)^2
+\left(\frac{\partial\varTheta}{\partial v}\right)^2
\biggr]+\frac{p_\varphi^2}{a^2\sin^2\!u\sinh^2\!v}\nonumber\\
&\quad=2m[E-U(u)].
\end{align}
It is expected in the tunnelling region that there does not exist oscillating-wave mode in the $u$-direction nor damping-wave mode in the $v$-direction (along an equipotential curve).
It might be hence reasonable to assume that
\begin{subequations}
\begin{equation}
\frac{\partial\varTheta_R}{\partial u}\simeq0,\quad
\frac{\partial\varTheta_I}{\partial v}\simeq0,
\label{eqn:Stationarity}
\end{equation}
and
\begin{equation}
\frac{a\sin u\sinh v}{\sqrt{g}}
\left(\frac{\partial\varTheta_I}{\partial u}\right)^2
\simeq2m[U(u)-W],
\label{eqn:ImaginaryEq}
\end{equation}
\begin{equation}
\frac{a\sin u\sinh v}{\sqrt{g}}
\left(\frac{\partial\varTheta_R}{\partial v}\right)^2
+\frac{p_\varphi^2}{a^2\sin^2\!u\sinh^2\!v}
\simeq2m(E-W),
\label{eqn:RealEq}
\end{equation}
\end{subequations}
where $W$ ($<E<U$) is regarded as the normal energy.
We obtain from Eqs.~(\ref{eqn:ImaginaryEq}) and~(\ref{eqn:RealEq})
\begin{subequations}
\label{eqn:Solution}
\begin{equation}
\varTheta_I(u,v)\simeq\rint ds\,\sqrt{2m[U(u)-W]},
\end{equation}
\begin{equation}
\varTheta_R(u,v)
\simeq\rint d\ell\,\sqrt{2m(E-W)-\frac{p_\varphi^2}{a^2\sin^2\!u\sinh^2\!v}}.
\end{equation}
\end{subequations}
$ds$ and $d\ell$ are the length elements along the curve $v$, $\varphi=\text{const.}$~and the curve $u$, $\varphi=\text{const.}$, respectively, given by~\cite{note:LengthElement} 
\begin{subequations}
\begin{equation}
ds=a\sqrt{\sin^2\!u\cosh^2\!v+\cos^2\!u\sinh^2\!v}\,du,
\label{eqn:LengthElement}
\end{equation}
\begin{equation}
d\ell=a\sqrt{\sin^2\!u\cosh^2\!v+\cos^2\!u\sinh^2\!v}\,dv.
\label{eqn:LengthElementV}
\end{equation}
\end{subequations}
One can confirm the consistency of the solution~(\ref{eqn:Solution}) with Eq.~(\ref{eqn:Stationarity}) under the situation of small curvature and short tunnelling length mentioned in the previous paragraph.
The barrier penetration probability $D(W,v)$ for an electron at $(v,\varphi)$ is thus given by
\begin{subequations}
\label{eqn:WKB}
\begin{equation}
D(W,v)=e^{-2S(W,v)/\hbar},
\end{equation}
\begin{equation}
S(W,v)=\rint_{u_0}^{u_T}ds\,\sqrt{2m[U(u)-W]}.
\label{eqn:WKBAction}
\end{equation}
\end{subequations}
Note that $u_T$ is determined by the equation $U(u_T)-W=0$.

The supply function at each point on the surface of the potential barrier is given by the same formula as that for the planar emitter model, which is derived in Ref.~\onlinecite{ref:Young}, since the surface of the barrier seems planar to electrons around that point.
Hence the supply function $N(W,E,v)\,dv\,d\varphi$ reads
\begin{equation}
N(W,E,v)\,dv\,d\varphi
=\frac{m}{2\pi^2\hbar^3}\frac{1}{e^{(E+\phi)/k_BT}+1}\,d^2\sigma,
\end{equation}
where $d^2\sigma$ is the areal element on the surface of the emitter defined by
\begin{equation}
d^2\sigma=\rho\,d\varphi\,d\ell\quad\text{with}\quad u=u_0.
\label{eqn:ArealElement}
\end{equation}
$\rho$ is given in Eq.~(\ref{eqn:EllipsoidalCordinate}) and $d\ell$ in Eq.~(\ref{eqn:LengthElementV}).

With these components together, we first calculate the (total) energy distribution of emitted electrons~\cite{ref:Young}
\begin{equation}
P(E)
=\rint_0^{2\pi}d\varphi\rint_0^\infty dv\rint_{-\infty}^EdW
N(W,E,v)D(W,v),
\label{eqn:EnergyDistributionIntegral}
\end{equation}
and then obtain the emission current $I$ by
\begin{equation}
I=e\rint_{-\infty}^\infty dE\,P(E).
\label{eqn:CurrentIntegral}
\end{equation}

Before going on any further, let us summarize the differences between our formulation and that of He \textit{et~al.}~in Ref.~\onlinecite{ref:Cutler}.
(i)~We neglect the image charge potential while He \textit{et~al.}~calculated it exactly.\cite{ref:CutlerImage}
(ii)~We count the electrons emitted from all over the surface of the emitter through the $v$-integration in Eq.~(\ref{eqn:FNEq}) while He \textit{et~al.}~assumed, as in the ordinary F--N theory, that the emission area on the surface of the emitter is small enough and is regarded as a plane.
These two points will be addressed in the following.

\subsection{Energy distribution $P(E)$}
Unfortunately, it is not possible to carry out the integrations exactly, but a few reasonable approximations enable us to achieve analytical formulae.
The first one is the linear approximation of the potential $V(u)$ in Eq.~(\ref{eqn:ElectricPotential}), which one may realize from Fig.~\ref{fig:Potential}(b):
\begin{equation}
V(u)
\simeq\frac{V_0}{\sin u_0\ln\cot(u_0/2)}(u-u_0).
\label{eqn:LinearApproximation}
\end{equation}
This is valid if $(u_T-u_0)\cot u_0\simeq(-W/eV_0)\cos u_0\times\ln\cot(u_0/2)\ll1$, which is satisfied in conventional experiments: $(u_T-u_0)\cot u_0\sim10^{-2}$ for $R\sim100\,\nm$, $L\sim5\,\cm$, $V_0\sim3\,\kV$, and $W\sim-\phi\sim-4.4\,\eV$.
(In this section, we will often demonstrate the validity of approximations with this set of parameters. We hereafter call it ``case I'' for short. Of course, this is not the only situation to which our formulae are applicable.)
In this regime, one can evaluate the action $S(W,v)$, defined in Eq.~(\ref{eqn:WKBAction}), for $v/\sin u_0\ll1$ (near the apex of the emitter) as
\begin{align}
\lefteqn{S(W,v)
\simeq\sqrt{\frac{2meV_0}{\sin u_0\ln\cot(u_0/2)}}
\rint_{u_0}^{u_T}ds\,\sqrt{u_T-u}}\nonumber\\
&\qquad\simeq S_0(W)\,\biggl[
1+\frac{1}{2}\left(\frac{v}{\sin u_0}\right)^2\nonumber\\
&\phantom{\qquad\simeq S_0(W)\,\biggl[}
{}-\frac{1}{8}\left(1-\frac{4}{3}\sin^2\!u_0\right)
\left(\frac{v}{\sin u_0}\right)^4
+\cdots
\biggr],
\label{eqn:Action}
\end{align}
where the factor $S_0(W)=S(W,0)$ is the action along the $z$-axis given by
\begin{equation}
S_0(W)
\simeq\frac{2a\sqrt{2m(-W)^3}}{3eV_0}\sin^2\!u_0\ln\cot(u_0/2).
\label{eqn:TipAction}
\end{equation}
In this evaluation, we have expanded the length element $ds$ given in Eq.~(\ref{eqn:LengthElement}) around $u=u_0$ and $v=0$ as
\begin{widetext}
\begin{equation}
\begin{split}
ds\simeq a\sin u_0\,\biggl[&
\Bigl(1+(u-u_0)\cot u_0\Bigr)
+\frac{1}{2}\,\Bigl(1-(u-u_0)\cot u_0\Bigr)
\left(\frac{v}{\sin u_0}\right)^2\\ 
&{}-\frac{1}{8}\left(
1-\frac{4}{3}\sin^2\!u_0+\frac{4}{3}(u-u_0)\sin u_0\cos u_0
\right)\left(\frac{v}{\sin u_0}\right)^4
+\cdots
\biggr]\,du,
\end{split}
\end{equation}
\end{widetext}
and neglected $O\bm{(}(u_T-u_0)\cot u_0\bm{)}$ terms.
The term $(v/\sin u_0)^4$ in Eq.~(\ref{eqn:Action}) is negligible in the formu\-la~(\ref{eqn:EnergyDistributionIntegral}) if $(v/\sin u_0)^4/8\ll(v/\sin u_0)^2/2$ is satisfied within the region $v<\sin u_0/\sqrt{2S_0(W)/\hbar}$ [which is the region where the barrier penetration probability $D(W,v)$ in Eq.~(\ref{eqn:WKB}) is not zero], i.e., $8S_0(W)/\hbar\gg1$.
Note here that the WKB approximation~(\ref{eqn:WKB}) is a semi-classical approximation and is valid for large $S(W,v)/\hbar$.
Typical value of $S_0(W)/\hbar$ is $S_0(-\phi)/\hbar\sim7.6$ in the case I\@.
The areal element $d^2\sigma$ in Eq.~(\ref{eqn:ArealElement}) is also approximated by
\begin{equation}
d^2\sigma\simeq a^2v\sin^2\!u_0\,dv\,d\varphi
\end{equation}
for $v/\sin u_0\ll1$, and the integration one should carry out is now
\begin{equation}
\begin{split}
\lefteqn{P(E)\simeq\frac{ma^2\sin^2\!u_0}{\pi\hbar^3}
\frac{1}{e^{(E+\phi)/k_BT}+1}}\\ 
&{}\times\rint_{-\infty}^EdW
\rint_0^\infty dv\,v\exp\biggl[
-\frac{2}{\hbar}S_0(W)
\,\biggl\{1+\frac{1}{2}\left(\frac{v}{\sin u_0}\right)^2\biggr\}
\biggr],
\end{split}
\label{eqn:AppIntegral}
\end{equation}
where the action $S_0(W)$ is given in Eq.~(\ref{eqn:TipAction}).
We thus reach the energy distribution formula for the hy\-perboloidal emitter model:
\begin{equation}
\begin{split}
P(E)={}&\frac{2m\phi LR^2}{3\pi\hbar^3(L+R)}
\frac{1}{e^{(E+\phi)/k_BT}+1}\\ 
&{}\times\frac{1}{(2S_0/\hbar)^{2/3}}
\varGamma\bm{(}-1/3,2S_0(E)/\hbar\bm{)},
\end{split}
\label{eqn:EnergyDistributionDim}
\end{equation}
where
\begin{equation}
S_0=S_0(-\phi)
=\frac{2L\sqrt{2m\phi^3}}{3eV_0\beta_0}
\label{eqn:CharAction}
\end{equation}
with a modification factor for field strength,
\begin{equation}
\beta_0=\frac{\cos u_0}{\sin^2\!u_0\ln\cot(u_0/2)},
\label{eqn:EnhancementFactor}
\end{equation}
which comes from the sharpness of the emitter, and
\begin{equation}
\varGamma(z,p)=\rint_p^\infty dt\,t^{z-1}e^{-t}
\end{equation}
is the incomplete gamma function.
Since $\varGamma(z,p)$ has an asymptotic expansion
\begin{equation}
\begin{split}
\lefteqn{\varGamma(z,p)=p^{z-1}e^{-p}}\\ 
&{}\times\left[
1+\sum_{n=1}^{N-1}\frac{1}{p^n}(z-1)(z-2)\cdots(z-n)+O(|p|^{-N})
\right]
\end{split}
\label{eqn:AsymptoticExpansion}
\end{equation}
for large $|p|$,\cite{ref:IntegralTable} the energy distribution~(\ref{eqn:EnergyDistributionDim}) is further approximated by
\begin{equation}
\begin{split}
P(E)\simeq{}&\frac{2m\phi LR^2}{3\pi\hbar^3(L+R)}
\frac{1}{e^{(E+\phi)/k_BT}+1}\\ 
&{}\times\frac{1}{(2S_0/\hbar)^2}
\exp\!\left[
-(2S_0/\hbar)\left(1-\frac{3}{2}\frac{E+\phi}{\phi}\right)
\right]
\end{split}
\label{eqn:EnergyDistributionAsymp}
\end{equation}
when the action $2S_0/\hbar$ is large.
Note that we had already assumed $8S_0/\hbar\gg1$ in Eq.~(\ref{eqn:AppIntegral}).
Noticing the high energy cut-off by the Fermi--Dirac distribution and the low energy one by the exponential factor in the expansion~(\ref{eqn:AsymptoticExpansion}), the action $S_0(E)$ has been expanded around the Fermi level $E\sim E_F=-\phi$ in Eq.~(\ref{eqn:EnergyDistributionAsymp}).
The energy distributions~(\ref{eqn:EnergyDistributionDim}) and~(\ref{eqn:EnergyDistributionAsymp}) are derived under the conditions
\begin{subequations}
\begin{equation}
(u_T-u_0)\cot u_0
\simeq\frac{L}{R}\frac{\phi}{eV_0\beta_0}
\simeq\frac{3S_0}{2R\sqrt{2m\phi}}
\ll1
\end{equation}
and
\begin{equation}
8S_0/\hbar\gg1,
\end{equation}
\end{subequations}
namely, they are valid for the parameters satisfying
\begin{equation}
1\ll8S_0/\hbar\ll\frac{16R\sqrt{2m\phi}}{3\hbar}.
\label{eqn:ParameterRegion}
\end{equation}

\begin{figure}[b]
\includegraphics[width=0.40\textwidth]{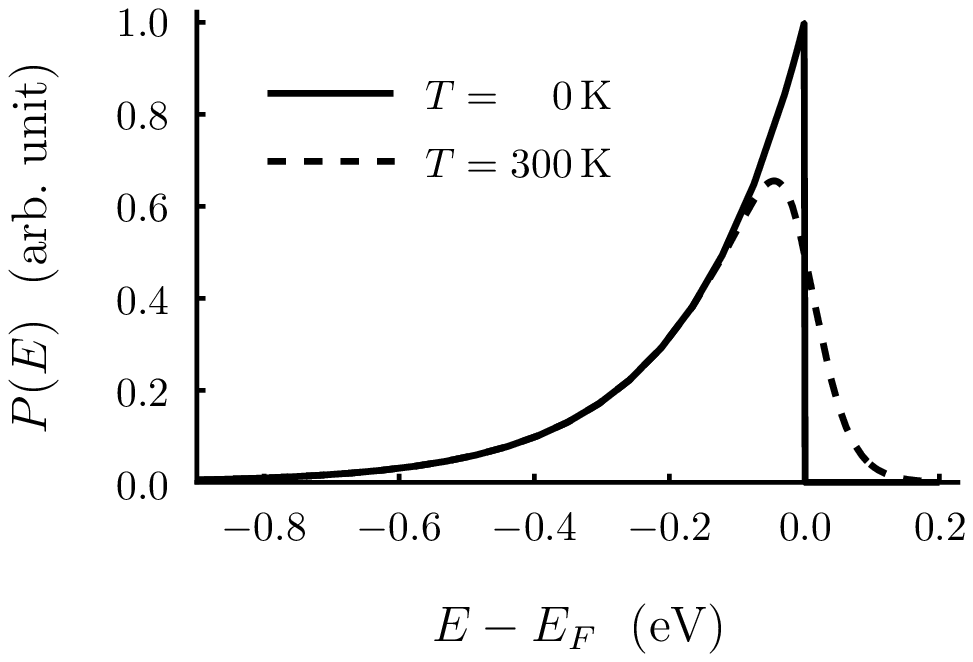}\\
\begin{picture}(0,0)
\put(-27,83){\makebox(0,0){\large(a)}}
\end{picture}
\bigskip\\
\includegraphics[width=0.40\textwidth]{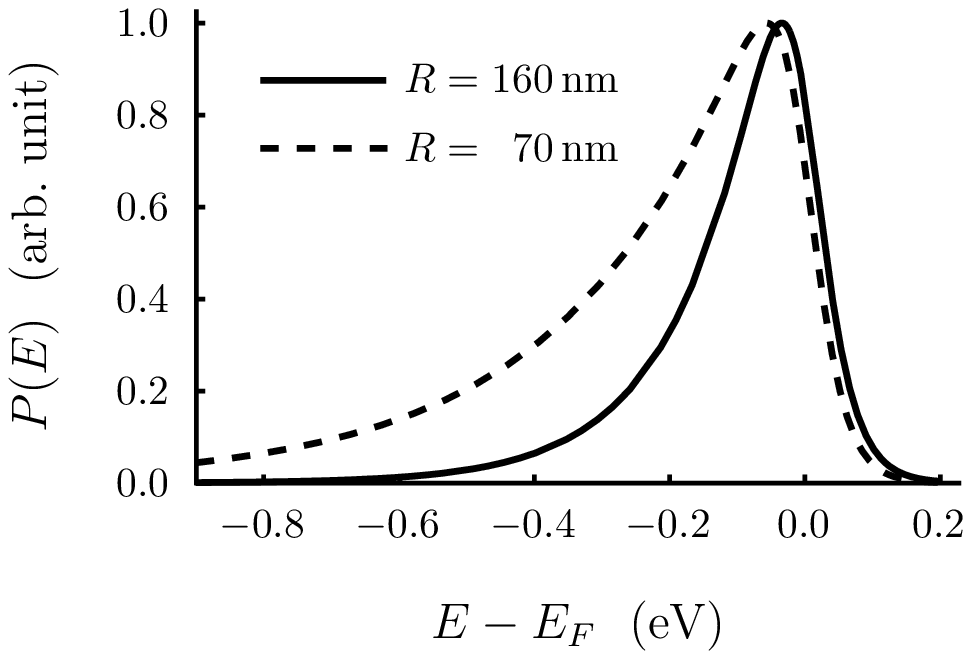}\\
\begin{picture}(0,0)
\put(-27,88){\makebox(0,0){\large(b)}}
\end{picture}
\caption{(a)~The energy distributions~(\ref{eqn:EnergyDistributionAsymp}) for zero and nonzero temperatures. The radius of curvature of the emitter is $R=110\,\nm$ ($u_0=0.0015$). The peak of the distribution for zero temperature is normalized to unity. (b)~The energy distributions~(\ref{eqn:EnergyDistributionAsymp}) for different radii of curvature. $u_0=0.0018$ for $R=160\,\nm$ and $u_0=0.0012$ for $R=70\,\nm$. Temperature is $T=300\,\K$. The peak of each distribution is normalized to unity. Other parameters for both figures~(a) and~(b) are those of the case I\@.}
\label{fig:EnergyDistribution}
\end{figure}
The energy distribution~(\ref{eqn:EnergyDistributionAsymp}) is plotted in Fig.~\ref{fig:EnergyDistribution}.
It is clear from the expression~(\ref{eqn:EnergyDistributionDim}) and~(\ref{eqn:EnergyDistributionAsymp}) that the action $S_0$ as well as the temperature $T$ characterizes the energy distribution $P(E)$.
Roughly speaking, the temperature $T$ determines the high energy cut-off of the distribution [Fig.~\ref{fig:EnergyDistribution}(a)], and the action $S_0$ the low energy one [Fig.~\ref{fig:EnergyDistribution}(b)].
For a sharper emitter, the action $S_0$ is smaller [see Eqs.~(\ref{eqn:CharAction}) and~(\ref{eqn:EnhancementFactor})], and the energy distribution $P(E)$ is wider [Fig.~\ref{fig:EnergyDistribution}(b)].
This agrees with and supports the numerical calculation by He \textit{et~al.}~in Ref.~\onlinecite{ref:Cutler}.
Furthermore, we have an explicit expression for a measure of the width of the distribution, $\varDelta E$:
\begin{equation}
\varDelta E
=\frac{\pi k_BT}{\sin(3\pi k_BTS_0/\hbar\phi)},
\label{eqn:Width}
\end{equation}
which is the deviation of the distribution $P(E)$ defined by $\varDelta E=\sqrt{\langle E^2\rangle-\langle E\rangle^2}$ where $\langle E^n\rangle=\rint_{-\infty}^\infty dE\,E^nP(E)/\break\rint_{-\infty}^\infty dE\,P(E)$.
Remember the integral $\rint_{-\infty}^\infty dx\,e^{x/d}/\break(e^{x/k_BT}+1)=\pi k_BT/\sin(\pi k_BT/d)$ for $k_BT<d$.\cite{ref:IntegralTable}
One can see from the formula~(\ref{eqn:Width}) that $\varDelta E$ is more sensitive to the action $S_0$ than to the temperature $T$.
In fact, the factor $(3\pi k_BTS_0/\hbar\phi)/\sin(3\pi k_BTS_0/\hbar\phi)$ is almost $1$ (about $1.03$ at room temperature $T\sim300\,\K$ in the case I), and $\varDelta E$ is approximated by $\varDelta E\simeq\hbar\phi/3S_0$.

\subsection{Emission area}
\begin{figure}[b]
\includegraphics[width=0.18\textwidth]{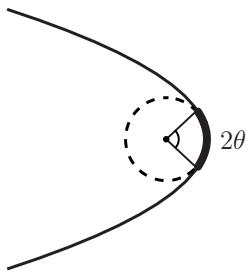}
\caption{Emission area measured by the angle $2\theta$ seen from the center of curvature.}
\label{fig:PinTopAngle}
\end{figure}

At this point, we can discuss the assumption on the emission area in the ordinary F--N theory and the theory of He \textit{et~al.},\cite{ref:Cutler} where the emission areas are assumed to be small enough and are regarded as planes.
In our theory, the emission area is not given by hand but is determined by the barrier penetration probability $D(W,v)$:
It is the region where $D(W,v)$, which is the exponential factor in the $v$-integration in Eq.~(\ref{eqn:AppIntegral}), is not zero and is estimated as the region $v<\sin u_0/\sqrt{2S_0/\hbar}$.
This seems at first sight to mean that the emission area is smaller for a sharper emitter and to support the above assumption.
This is however not true.
One should take the curvature of the area into account.
The relevant measure is not the size itself but the solid angle of the area seen from the center of curvature, i.e., the angle $2\theta$ in Fig.~\ref{fig:PinTopAngle}.
The angle $\theta$ is evaluated as
\begin{equation}
\tan\theta
=\frac{\sin u_0\cos u_0\sinh(\sin u_0/\sqrt{2S_0/\hbar})}{1-\cos^2\!u_0\cosh(\sin u_0/\sqrt{2S_0/\hbar})}
\simeq\frac{\sqrt{\hbar/2S_0}}{1-\hbar/4S_0}
\label{eqn:SolidAngle}
\end{equation}
(for small $u_0$), which shows that, for a sharper emitter, the action $S_0$ is smaller, and the angle $2\theta$ is larger.
In this sense, the emission area for a sharp emitter is not planar.
Even in a conventional situation as the case I, the action is $S_0/\hbar\sim7.6$ and the angle is $2\theta\sim30^\circ$, which is not so small that the emission area can be regarded as planar.

\subsection{Field emission current $I$}
It is also possible to execute the final integration~(\ref{eqn:CurrentIntegral}) with the expression~(\ref{eqn:EnergyDistributionDim}) if temperature $T=0$.
The emission current $I$ at zero temperature is then
\begin{equation}
\begin{split}
I={}&\frac{2me\phi^2LR^2}{3\pi\hbar^3(L+R)}
\frac{1}{(2S_0/\hbar)^{4/3}}\\
&{}\times\Bigl(
\varGamma(1/3,2S_0/\hbar)
-(2S_0/\hbar)^{2/3}\varGamma(-1/3,2S_0/\hbar)
\Bigr).
\end{split}
\label{eqn:CurrentDim}
\end{equation}
If the action $2S_0/\hbar$ is large, however, a current formula for finite temperature is available:
Integrating the energy distribution~(\ref{eqn:EnergyDistributionAsymp}), one obtains
\begin{equation}
I=\frac{4me\phi^2LR^2}{9\pi\hbar^3(L+R)}
\frac{3\pi k_BTS_0/\hbar\phi}{\sin(3\pi k_BTS_0/\hbar\phi)}\frac{1}{(2S_0/\hbar)^3}e^{-2S_0/\hbar}.
\label{eqn:CurrentAsymp}
\end{equation}
These current formulae are again valid under the condition~(\ref{eqn:ParameterRegion}).
It has been known experimentally and the F--N theory has explained successfully that field emission current does not depend on temperature significantly.\cite{ref:FowlerNordheim,ref:GoodMueller}
This is also the case with our hyperboloidal emitter model.
As already mentioned in the previous subsection, the factor $3\pi k_BTS_0/\hbar\phi$ found in Eq.~(\ref{eqn:CurrentAsymp}) is small and the temperature correction to the current is only about $3\%$ at room temperature $T\sim300\,\K$\@.
The temperature dependence is hence negligible in the emission current $I$\@.

Since the action $S_0$ is proportional to $V_0^{-1}$ [see Eq.\ (\ref{eqn:CharAction})], the emission current~(\ref{eqn:CurrentAsymp}) offers the following current--voltage characteristics for the hyperboloidal emitter:
\begin{subequations}
\label{eqn:I-V}
\begin{equation}
\frac{I}{V_0^3}\propto\exp\!\left(-\frac{A}{V_0}\right),
\label{eqn:I-VMain}
\end{equation}
\begin{equation}
A=\frac{4L\sqrt{2m\phi^3}}{3\hbar e\beta_0},\quad
\beta_0=\frac{\cos u_0}{\sin^2\!u_0\ln\cot(u_0/2)}.
\end{equation}
\end{subequations}
Remember the corresponding relationship in the F--N theory, $I/V_0^2\propto\exp(-A/V_0)$.\cite{ref:FowlerNordheim,ref:Nordheim,ref:GoodMueller,ref:Young}
Taking the geometry of emitter into account, we have obtained different exponent of $V_0$ on the left-hand side.
This is one of the main results of this article.
The new exponent is a direct consequence of the Gaussian integral over the variable $v$ in Eq.~(\ref{eqn:AppIntegral}), or roughly speaking, of the finiteness of the emission area, which comes from the finiteness of the radius of curvature of the emitter.
It is easily expected that the similar situations arise for various shapes of emitter, where the integration over a finite emission area yields some exponent [not necessarily the same as in the formula~(\ref{eqn:I-V}) but] different from the conventional one.
The geometrical effect, in general, manifests itself in the exponent.\cite{note:Hyper2Planar} 

It is interesting, on the other hand, that the exponential factor on the right-hand side of Eq.~(\ref{eqn:I-VMain}) is the same as that of the conventional one.
This, in some sense, supports the validity of the planar model, since the current--voltage characteristics are mainly dominated by this exponential factor but the power on the left-hand side is relatively less important.
It is not easy to observe the difference in the exponent of $V_0$ experimentally,\cite{ref:GoodMueller} and the conventional F--N theory works well \textit{fortunately}.

Our formula~(\ref{eqn:I-V}) however does not coincide with the one that He \textit{et~al.}~concluded in Ref.~\onlinecite{ref:Cutler} for the hyperboloidal emitter model, i.e., $I/V_0^2\propto\exp(-A/V_0-B/V_0^2)$.
What is the reason for this discrepancy or what is the origin of the term $-B/V_0^2$?
The discrepancy seems at first sight due to the fact that the image charge effect is neglected in our formulation while He \textit{et~al.}~calculated it exactly.
This is not the case, however.
In general, the image charge effect becomes more prominent as the applied voltage increases, and the emission current is accordingly enhanced.
But such an effect is not observed in their calculation.
The additional $-B/V_0^2$ term in the theory of He \textit{et~al.}~comes not from the image charge effect but from the bias field potential.
In fact, if one retains the term of the order of $(u_T-u_0)\cot u_0\simeq(L/R)(-W/eV_0\beta_0)$ in the action integral~(\ref{eqn:TipAction}), which we have neglected there, such a correction appears:
\begin{equation}
S_0(W)
=\frac{2L\sqrt{2m(-W)^3}}{3eV_0\beta_0}
\left(1+\alpha\frac{L}{R}\frac{(-W)}{eV_0\beta_0}+\cdots\right),
\label{eqn:CorrectedAction}
\end{equation}
where $\alpha$ is a positive constant of the order of $1$.\cite{note:PositiveCorrection} 
This correction comes in when the emitter is so sharp that the linear approximation of the bias field potential $V(u)$ in Eq.~(\ref{eqn:LinearApproximation}) is not valid.
Note the $R$-dependence of $(u_T-u_0)\cot u_0\simeq3S_0(W)/2R\sqrt{2m(-W)}$.
It should be noted, however, that emitters with very sharp radii do not fit in well with our formulation as mentioned in the second paragraph of this section.

\section{Estimation of Experiments}
\label{eqn:Experiment}
In our formulae, we have the parameter $u_0$, which characterizes the sharpness of the hyperboloidal emitter and is not contained in the ordinary F--N theory based on the planar emitter model.
Finally, let us try to analyze actual experiments by making use of this feature.
The experimental data to be fitted in the following were obtained for tungsten emitters in the experiments whose details are presented in Ref.~\onlinecite{ref:NanotipOshimaASS182}.
In those experiments, the parameter $L$ is about $L\sim5\,\cm$.

\subsection{Current--voltage characteristics}
\begin{figure}[b]
\includegraphics[width=0.40\textwidth]{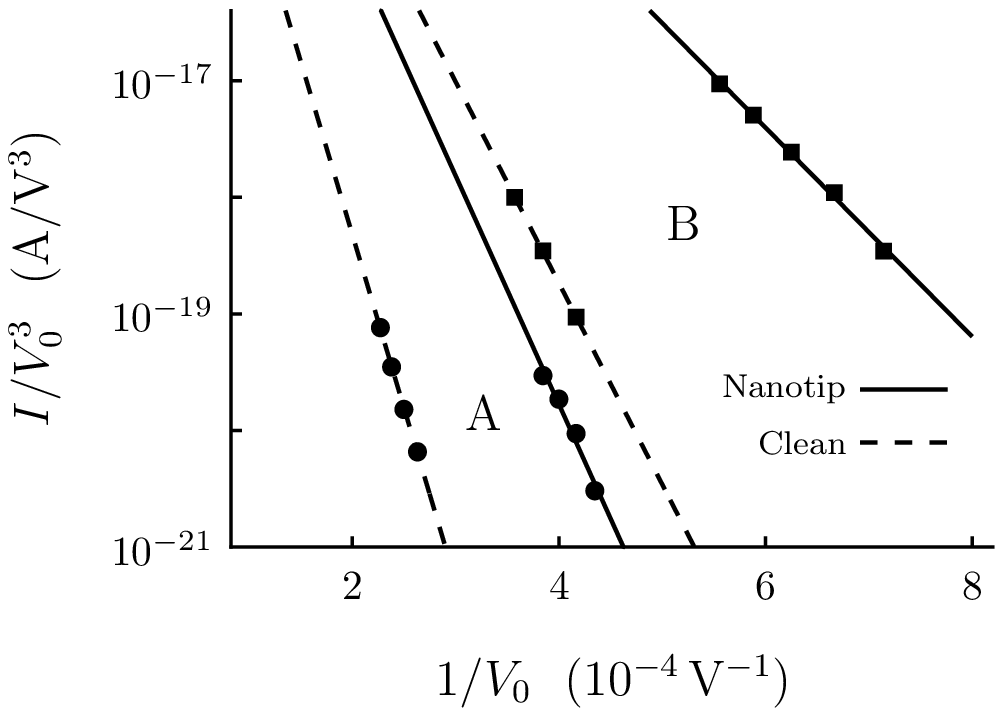}\bigskip\\
\includegraphics[width=0.40\textwidth]{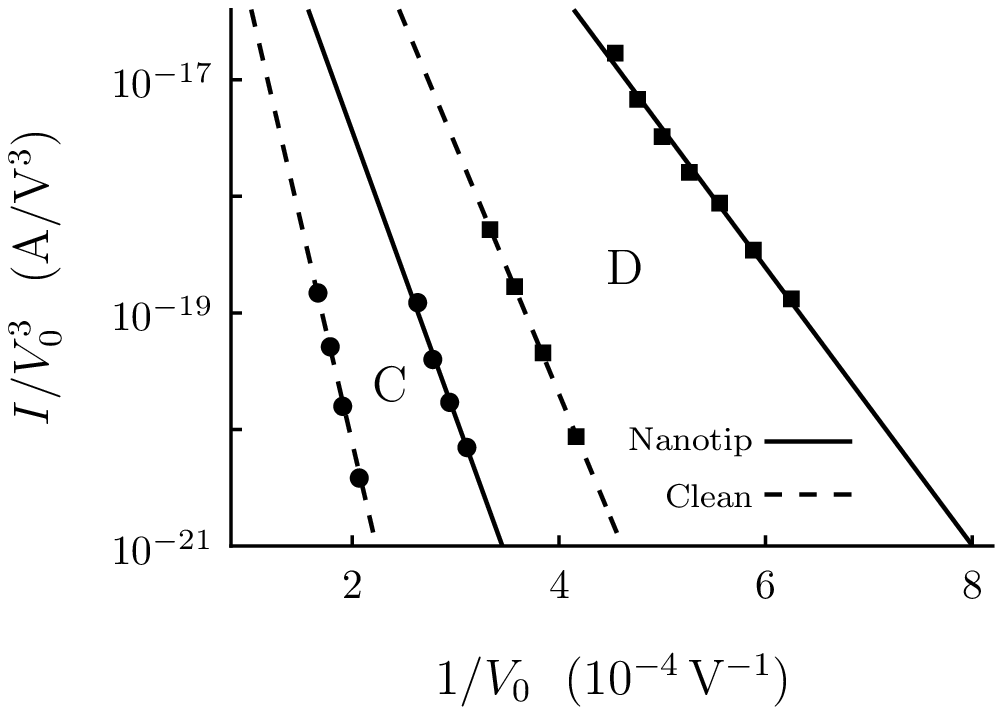}
\caption{Experimental data of current--voltage characteristics for four different pairs (A--D) of emitters: Dashed lines are for normal clean emitters, and solid lines for nanotips fabricated on top of the clean ones.\cite{ref:NanotipOshimaASS182} Each series of data is fitted by the formula~(\ref{eqn:I-V}).}
\label{fig:FNFits}
\end{figure}

Experimental data of current--voltage characteristics for different emitters A--D are shown in Fig.~\ref{fig:FNFits}.
Not only those for normal clean emitters but also those for nanotip emitters~\cite{ref:NanotipOshimaASS182} which are fabricated on the apexes of the formers are plotted in the same figures.
The vertical axis $I/V_0^3$ in Fig.~\ref{fig:FNFits} is different from that of the ordinary F--N plot, $I/V_0^2$, and each series of data is well fitted by a straight line, i.e., by the formula~(\ref{eqn:I-V}).

\begin{figure}
\begin{tabular}{c@{\qquad\qquad}c}
\includegraphics[height=0.12\textwidth]{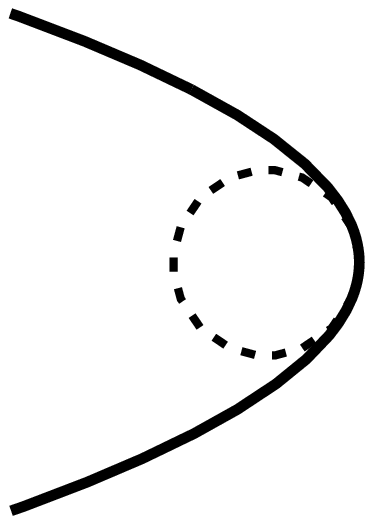}&
\includegraphics[height=0.12\textwidth]{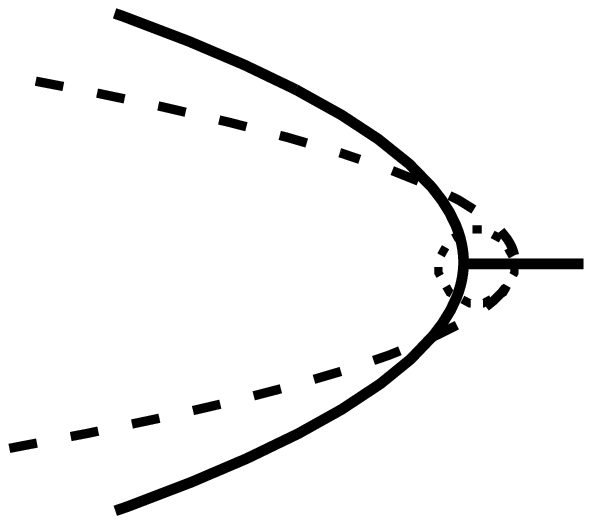}\\
(a)&(b)
\end{tabular}
\caption{(a)~Radius of curvature of a clean emitter and (b)~effective radius of curvature of a nanotip emitter.}
\label{fig:R}
\end{figure}

As already mentioned in the previous section, fittings of the data with the conventional F--N formula also work well similarly to the ones with our formula shown in Fig.~\ref{fig:FNFits}\@.
Our formula~(\ref{eqn:I-V}) however contains the parameter $u_0$, which characterizes the sharpness of the emitter, while the conventional one does not.
We can hence estimate the radii of curvature $R$ of the emitters by the fittings, which are listed in Table~\ref{tab:RI-V}\@.
\begin{table}[t]
\catcode`?=\active \def?{\phantom{0}}
\caption{Radii of curvature of the emitters estimated by the fittings in Fig.~\ref{fig:FNFits} with $\phi=4.4\,\eV$ and $L=5\,\cm$.}
\begin{ruledtabular}
\begin{tabular}{ccc}
 &\multicolumn{2}{c}{Radius of curvature $R$ ($\nm$)}\\
 &Clean&Nanotip\\\hline
A&$154$&$?99$\\
B&$?86$&$?43$\\
C&$201$&$125$\\
D&$108$&$?58$
\end{tabular}
\end{ruledtabular}
\label{tab:RI-V}
\end{table}
It should be noted there that the radii of curvature for the nanotips are not exactly those of the \textit{tips} of the emitters:
They are just the radii of curvature of \textit{effective} clean emitters which emit the same currents as the actual emitters with nanotips (Fig.~\ref{fig:R}).
One can see from this estimation, however, that a nanotip fabricated on top of a clean emitter has such an effect that it reduces the radius of curvature of the emitter to the extent of a half of the original clean one.

\subsection{Energy distribution}
\begin{figure}[t]
\includegraphics[width=0.40\textwidth]{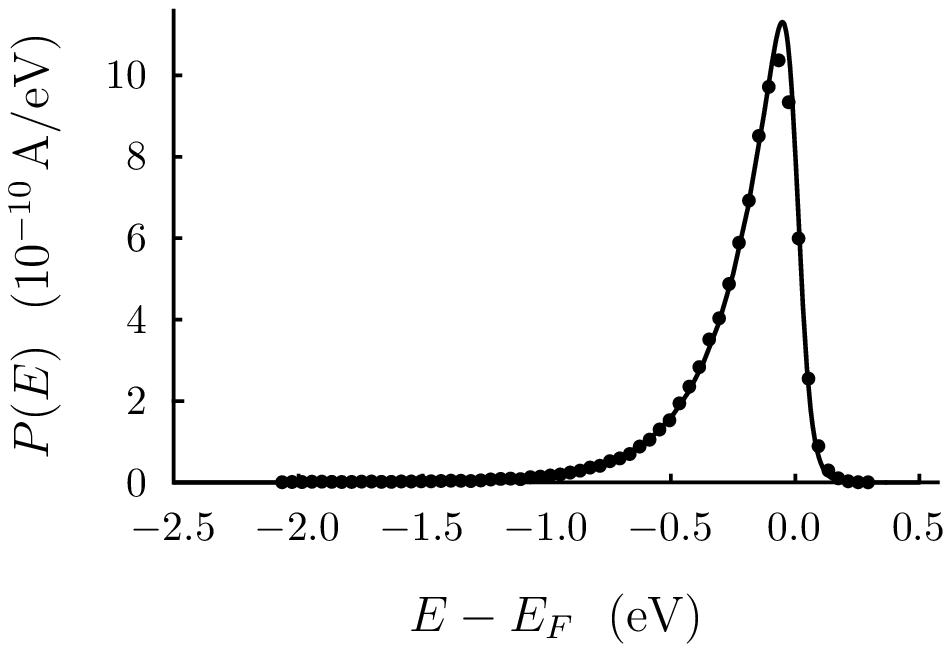}\\
\begin{picture}(0,0)
\put(-20,133){\makebox(0,0){\large(a)}}
\end{picture}
\bigskip\\
\includegraphics[width=0.40\textwidth]{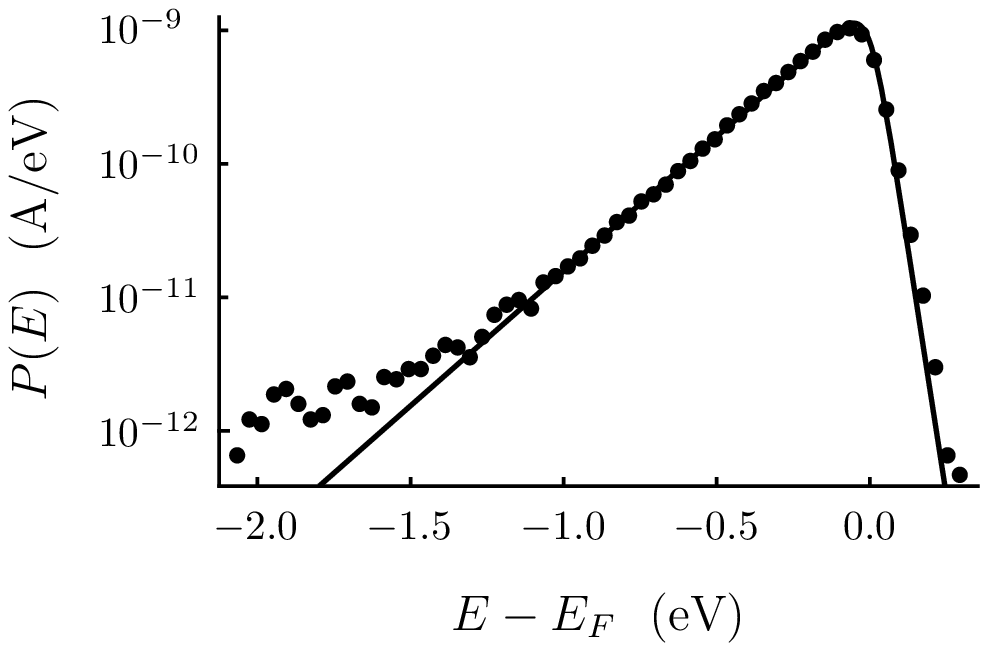}\\
\begin{picture}(0,0)
\put(-20,128){\makebox(0,0){\large(b)}}
\end{picture}
\caption{Experimental data (dots) of the energy distribution for the clean emitter A in Fig.~\ref{fig:FNFits} at $V_0=3.8\,\kV$, fitted (lines) by the formula~(\ref{eqn:EnergyDistributionAsymp}) in the log-linear plot~(b) with the parameters $E_F=-\phi=-4.4\,\eV$ and $T=300\,\K$ fixed. The linear one~(a) is drawn with the same parameters as in~(b).}
\label{fig:EneDisFits}
\end{figure}

Energy distributions for normal clean emitters are also fitted well by the formula~(\ref{eqn:EnergyDistributionDim}) and~(\ref{eqn:EnergyDistributionAsymp}) as shown in Fig.~\ref{fig:EneDisFits}.
The parameter $u_0$, i.e., the radius of curvature $R$ of emitter, is again obtained by the fitting.
The radius of curvature of the clean emitter A in Fig.~\ref{fig:FNFits}, which is estimated with the data for different applied voltages $V_0$, is presented in Table~\ref{tab:R-V}\@.
\begin{table}[b]
\caption{The radius of curvature $R$ of the clean emitter A in Fig.~\ref{fig:FNFits} estimated by fitting energy distributions for different applied voltages $V_0$ with the formula~(\ref{eqn:EnergyDistributionAsymp}). The parameters are fixed $\phi=4.4\,\eV$, $T=300\,\K$, and $L=5\,\cm$.}
\begin{ruledtabular}
\begin{tabular}{ccccc}
$V_0$ ($\kV$)&3.8&4.0&4.2&4.4\\\hline
$R$ ($\nm$)  &114&113&111& 95
\end{tabular}
\end{ruledtabular}
\label{tab:R-V}
\end{table}
Although the radius of curvature $R$ should be independent of the applied voltage $V_0$, of course, the estimated value decreases slightly with $V_0$.
This is due to the image charge effect.
This effect is neglected in our formulation, and the enhancement of the emission current due to it is renormalized into the parameter $u_0$, i.e., into the sharpness of the emitter.
The enhancement is a slowly increasing function of $V_0$ as already mentioned in the previous section, and the estimated radius of curvature of the emitter accordingly decreases with $V_0$.
One hence has to be careful about the fact that the image charge effect is included in the parameter $u_0$.

\section{Conclusion}
In this article, we have derived field emission formulae from a \textit{hyperboloidal} emitter model, i.e., current--voltage characteristics~(\ref{eqn:CurrentDim}),~(\ref{eqn:CurrentAsymp}), and~(\ref{eqn:I-V}), and energy distribution of emitted electrons~(\ref{eqn:EnergyDistributionDim}) and~(\ref{eqn:EnergyDistributionAsymp}), which are valid under the condition~(\ref{eqn:ParameterRegion}).
Reflecting the geometry of the emitter, the traditional F--N formulae, which are derived based on the \textit{planar} emitter model, are modified.
The current--voltage characteristics in the F--N theory, $I/V^2\propto\exp(-A/V)$, is replaced with $I/V^3\propto\exp(-A/V)$, for example.
The geometrical effect manifests itself in the exponent of $V_0$ on the left-hand side.

We have also addressed and reconsidered the assumption of the planar emitter in the F--N theory.
An estimation of the spread of emission area based on the formula~(\ref{eqn:SolidAngle}) shows that the area cannot be regarded as planar even for a conventional emitter.

Furthermore, our analytical calculation has revealed the backgrounds of the conclusions drawn by He \textit{et~al.},\cite{ref:Cutler} which are based on a numerical calculation: The origin of the correction term in their current--voltage characteristics has been clarified, and the dependence of the width of energy distribution on radius of curvature of emitter has been explained.
The concise formula~(\ref{eqn:Width}) for the width of energy distribution might be useful in practical experiments.

And finally, we have attempted to analyze experimental data of nanotip emitters~\cite{ref:NanotipOshimaASS182} by making use of the parameter $u_0$, which characterizes the sharpness of the emitters, and clarified an effect of a nanotip fabricated on top of a normal clean emitter:
The effective radius of curvature of the emitter is considerably reduced.

One should note, however, that our formulae do not explain the characteristics peculiar to nanotip emitters exactly.
Actually, energy distributions for nanotip emitters,\cite{ref:NanotipBinhPRL69,ref:NanotipOshimaASS182} for example, cannot be fitted by our formula~(\ref{eqn:EnergyDistributionDim}) and~(\ref{eqn:EnergyDistributionAsymp}).
One of the reasons for this is that electrons cannot be treated as localized objects in nanotips, whose radii of curvature are of the order of $1\,\nm$.
For a rigorous description of field emission from today's nanoscale tips, a fully quantum treatment is required.\cite{ref:Lang,ref:NanotipGohdaPRL87}

\begin{acknowledgments}
This work was supported by Research for the Future Program in Japan Society for the Promotion of Science (JSPS), and the Grant-in-Aid for COE Research, MEXT\@.
\end{acknowledgments}


\begin{thebibliography}{16}
\bibitem{note:LengthElement}
The distance $dr$ between two points $(u,v,\varphi)$ and $(u+du,v+dv,\varphi+d\varphi)$ is given by $dr^2=a^2(\sin^2\!u\cosh^2\!v+\cos^2\!u\sinh^2\!v)(du^2+dv^2)+a^2\sin^2\!u\sinh^2\!v\,d\varphi^2$.
\bibitem{note:Hyper2Planar}
One may consider that our formula~(\ref{eqn:I-V}) should coincide with the conventional F--N formula in the limit $R\to\infty$. It is not the case, however, because the expansion of the action $S(W,v)$ in Eq.~(\ref{eqn:Action}) is valid only when the emission area is finite, but the area is infinite for conventional planar emitter model. The new formula~(\ref{eqn:I-V}), which is derived with this expansion, therefore, cannot be reduced to the conventional one even if the radius of curvature $R$ is taken to infinity.
\bibitem{note:PositiveCorrection}
It is easy to understand the sign of $-B/V_0^2$, i.e., positivity of $\alpha$ in Eq.~(\ref{eqn:CorrectedAction}), since $O\bm{(}(u_T-u_0)\cot u_0\bm{)}$ correction to the bias field potential $U(u)$ is always positive and gives rise to a positive correction to the action $S_0(W)$.
\end{thebibliography}

\end{document}